\documentclass[superscriptaddress,12pt]{revtex4}
\usepackage{amsfonts}
\usepackage{graphicx, amsmath}
\usepackage{subfigure}
\begin{document}
\title[Short Title]{High-efficiency alignment-free quantum cryptography based on quantum interference}
\author{Qi Guo}
\affiliation{Department of Physics, Harbin Institute of Technology, Harbin, Heilongjiang 150001, People's Republic of China}
\author{Liu-Yong Cheng}
\affiliation{Department of Physics, Harbin Institute of Technology, Harbin, Heilongjiang 150001, People's Republic of China}
\author{Hong-Fu Wang}
\affiliation{Department of Physics, College of Science, Yanbian University, Yanji, Jilin 133002, People's Republic of China}
\author{Shou Zhang\footnote{E-mail: szhang@ybu.edu.cn}}
\affiliation{Department of Physics, Harbin Institute of Technology, Harbin, Heilongjiang 150001, People's Republic of China}
\affiliation{Department of Physics, College of Science, Yanbian University, Yanji, Jilin 133002, People's Republic of China}
\begin{abstract}
\textbf{Abstract:}We propose an alternative quantum cryptography protocol using the quantum interference effect. The efficiency of creating sifted key can reach 100\% in principle, which is higher than previous protocols. Especially, compared with the typical quantum key distribution, the present scheme does not require the authorized parties to check their bases. Because the potential eavesdropper can only access part of the quantum system, the proposed scheme has natural practical security advantages. The scheme can be implemented with current technologies and opens promising possibilities for quantum cryptography.
\\\textbf{Keywords:}
Quantum cryptography, Quantum interference, Linear optics
\end{abstract}
\maketitle

Quantum cryptography provides an unconditional security method to distribute secret keys between two remote parties based on the fundamental principles of quantum mechanics \cite{1,2,3}. Quantum key distribution (QKD) is the first application of quantum mechanics in the practical communications. Due to the promising application prospect, quantum cryptography has been studied extensively. Various QKD protocols have been continuously proposed during the past two decades \cite{4,5,6}, and the experimental techniques of the practical QKD is reaching maturity \cite{7,8}. All these QKD protocols require signal particles to be transmitted in the quantum channel between the two parties, i.e. the creation of secret keys relies on the transmission of the carrier of secret information.

However, Noh presented a novel QKD protocol \cite{9}, called counterfactual quantum cryptography, which showed the QKD task can be accomplished even though signal particles were not transmitted through the quantum channel. Due to the counterfactuality and novelty, Noh's protocol has attracted widespread attention \cite{10,11,12,13,14,15}. The security of the counterfactual quantum cryptography was analyzed in an ideal situation \cite{11,12}, and the protocol was demonstrated in experiment \cite{13,14,15}. The basic idea of Noh's protocol originated from the interaction-free measurement \cite{16}. The presence of an obstacle in one of the arms of a Mach-Zehnder interferometer will destroy the interference even through no photon interacted with the obstacle. In Noh's protocol, it can be considered that Bob, one of the two participants, encoded secret bits by the presence or not of the obstacle. The creation of a sifted key is only from the events that interference is destroyed and the photon is not absorbed. The efficiency of the sifted key creation is 12.5\%, which has been improved to 50\% subsequently by Sun and Wen \cite{10}.

 Here we present an alternative quantum cryptography protocol just by using the quantum interference rather than destroying the interference. Compared with the existing protocols, the present scheme has higher efficiency of the sifted key creation that can be approached to 100\% in the ideal case. The security can be guaranteed by avoiding the eavesdropper to access the entire quantum system. The architecture of the proposed protocol is shown in Fig.~1. The two remote parties in the communication are called Alice and Bob. A long-armed Michelson-type interferometer is established between Alice and Bob. At the beginning Alice triggers the single-photon source SPS to emit a linearly polarized single photon. The polarization of the photon is chosen randomly, and Alice encodes horizontal polarization $|H\rangle$ and vertical polarization $|V\rangle$ as logic ``0" and ``1", respectively. The photon passes through the optical circulator C firstly, and is then rotated by the half-wave plate HWP oriented at 22.5$^\circ$, whose action can be given by \{$|H\rangle\rightarrow\frac{1}{\sqrt{2}}(|H\rangle+|V\rangle)$, $|V\rangle\rightarrow\frac{1}{\sqrt{2}}(|H\rangle-|V\rangle)$\}. Subsequently, the polarizing  beam splitter PBS$_1$ splits the photon into two pathes $a$ and $b$. The path $a$ leads to the mirror M$_1$ located in Alice's site, and the path $b$ goes toward the mirror M$_2$ at Bob's site. So after passing through the PBS$_1$, the quantum state can be expressed as
\begin{eqnarray}\label{e1}
&&|\phi_{0}\rangle=\frac{1}{\sqrt{2}}(|H\rangle_{b}+|V\rangle_{a}),
\cr&&|\phi_{1}\rangle=\frac{1}{\sqrt{2}}(|H\rangle_{b}-|V\rangle_{a}),
\end{eqnarray}
where subscripts 0 and 1 indicate Alice chooses bit value 0 and 1, respectively, i.e. she launches $|H\rangle$ and $|V\rangle$ photons. $a$ and $b$ label the two arms of the interferometer.

\begin{figure}
\scalebox{0.9}{\includegraphics{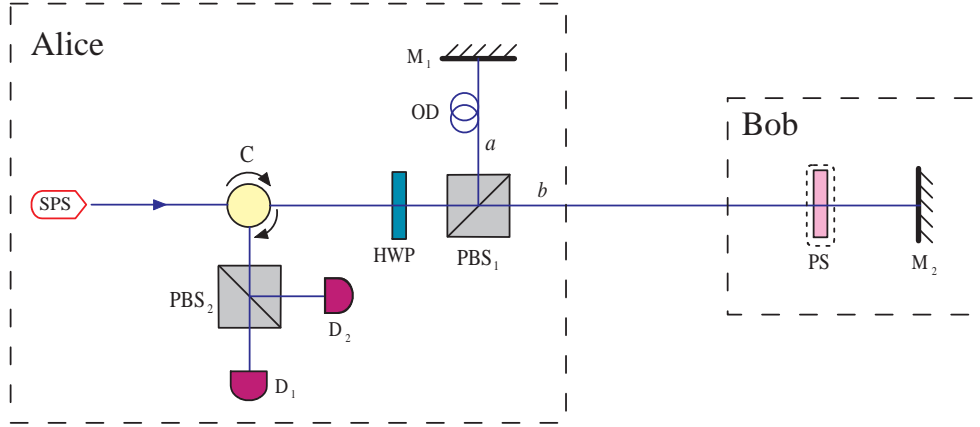}}\caption{\label{f1}
Schematic for the proposed QKD protocol. Two pathes $a$ and $b$ form a Michelson-type interferometer. SPS: single-photon source. C: optical circulator. HWP: half-wave plate. PBS: polarizing beam splitter. OD: optical delay used to maintain the phase by compensating the path difference of the two pathes. M: totally reflecting mirror. The phase shifter PS in a dashed box indicates that Bob can randomly remove or insert the PS in path $b$.}
\end{figure}

Phase shifter PS in Bob's site can introduce a $\pi$ phase for photons in the path $b$, i.e. $|H/V\rangle_{b}\rightarrow-|H/V\rangle_{b}$. Optical delay OD in path $a$ is used to match the optical path lengthes of the two arms. Bob encodes his bit on the presence or not of the PS, that is, removing the PS represents bit value ``0", and inserting the PS represents bit value ``1". Bob randomly modulates the phase of the photon pulse in path $b$ by \{0, $\pi$\}. Through straightforward calculation, it can be obtained that if Alice's and Bob's bit values are equal, the phase difference is 0 between the two paths, so after leaving the interferometer, the photon will go toward the detector $D_1$ by passing through C and PBS$_2$ with certainty owing to the interference effect. However, if Alice's and Bob's bit values are opposite, the phase difference of the two paths is $\pi$ and the photon will go toward the detector $D_2$. Therefore, if $D_1$ clicks, Alice and Bob can obtain an identical bit; if $D_2$ clicks, Alice flips her bit and they can also obtain an identical bit. Because of the loss or the interception of the potential eavesdropper Eve, it's possible that both $D_1$ and $D_2$ don't clicks. In this case, Alice announces her nondetection events, and they discard their bits. By this way, a sifted key can be established between Alice and Bob. In order to detect the potential eavesdropper Eve, they can select portions of the sifted key to estimate the error rate. Then they can also perform error correction and privacy amplification on the remaining sifted key as in conventional QKD protocol.

From the statement above, every Bob's detection result is effective for creating sifted key, which is different form the previous protocols with noneffective detection results. In ideal cases, every single-photon pulse sent in the interferometer can create an identical bit, the overall efficiency of creating a sifted key can thus reach 100\%. Therefore, the present protocol fundamentally improves the creation efficiency of sifted key. Furthermore, Alice only needs to announce the events that there is no detector clicks. Hence, the present scheme does not need that the authorized parties to check their bases during the process of creating sifted keys, which is very different from the other existing protocols, and classical communication resource is greatly saved.

The security of the present protocol can be verified by using the new type of no-cloning principle of orthogonal states proposed by Noh \cite{9}. The conventional no-cloning principle \cite{17,18} suggests that that orthogonal states in a composite system made of two subsystems cannot be cloned if the subsystems are only available one after the other. In Ref.~\cite{9}, Noh proposed a new type of no-cloning principle of orthogonal states for the case that only one subsystem can be accessed while the other one can never be accessed. That is, ``\emph{if reduced density matrices of an available subsystem are nonorthogonal and if the other subsystem is not allowed access, it is impossible to distinguish two orthogonal quantum states without disturbing them}". Adding the vacuum state in one of the arms of the interferometer, the quantum states after a photon enter the interferometer can be rewritten as
\begin{eqnarray}\label{e2}
&&|\Psi_{0}\rangle=\frac{1}{\sqrt{2}}(|0\rangle_{a}|H\rangle_{b}+|V\rangle_{a}|0\rangle_{b}),
\cr&&|\Psi_{1}\rangle=\frac{1}{\sqrt{2}}(|0\rangle_{a}|H\rangle_{b}-|V\rangle_{a}|0\rangle_{b}),
\end{eqnarray}
where $|0\rangle_{k}(k=a,b)$ denotes the vacuum state in the path $k$. The subsystem Eve can access is the path $b$, and the reduced density matrices of the available subsystem can be obtained $\mathrm{Tr}[\rho_{0}(\mathrm{path}~b)\rho_{1}(\mathrm{path}~b)]=\frac{1}{4}\neq0$, where $\rho_{i}(\mathrm{path}~b)=\mathrm{Tr}_{\mathrm{path}~ a}[|\Psi_{i}\rangle\langle\Psi_{i}|]~(i=0,1)$. Therefore, the reduced density matrices of the available subsystem path $b$ are nonorthogonal, and the other subsystem path $a$ is not allowed access for Eve. According to the new no-cloning principle of orthogonal states, Eve can't distinguish the two states without disturbing them, which ensures the security of the proposed protocol.

For the stabilized interferometer, a specific detector will be triggered for given Alice's and Bob's bit values due to the quantum interference. If Eve performs intercept-resend (I-R) attack on path $b$, the ``which-path" information of the photon will be revealed and the optical path length will also be prolonged, which are bound to destroy the interference and each of the two detectors may click with the probability of 50\% for all the bit values of Alice and Bob. That is, a 50\% error rate will be introduced in the sifted key. Thus, this attack will be discovered by the subsequent error estimation process. If the protocol is implemented by use of polarization multiphoton pulses, it may be undergo photon-number splitting (PNS) attack. However, the PNS attack will break the balance of the pulse intensity in the two pathes, and the two detectors may be triggered simultaneously, which will be found by Alice. Then she informs Bob to discard this bit. One the other hand, even through Eve intercept photons from path $b$, she cannot obtain any information about secret bit values of Alice and Bob. Therefore, the proposed protocol is robust against typical I-R attack and PNS attack strategies.

The experiment setup for implementing the proposed protocol is very simple, and all the components and devices are ordinary optical elements in optical laboratory. In the actual implementation, there may be two technical challenges. Firstly, the protocol is very dependent on quantum interference in the interferometer, so a stabilized long-armed interferometer should be established between Alice and Bob, which may be a challenge in experiment. The other challenge is that an ultrafast optical switch is needed for Bob to control the presence or absence of the PS in path $b$. Nevertheless, recent experimental progresses provide positive possibilities for overcoming these issues. Stabilized interferometers with several tens of kilometers have been reported \cite{19,20}. Suitable ultrafast optical switch with minimal loss and without disturbing the photon's quantum state has been demonstrated with switching window of 10 ps \cite{21,22}. Hence the present protocol is feasible under the current experimental conditions. In addition, it should be noted that there are two equivalent implementation schemes for the proposed protocol. The first one is that Alice uses the same encoding approach as Bob in path $a$. In this version, Alice only needs to launch single photons with identical polarization into the interferometer. The other equivalent approach is to  utilize a Mach-Zehnder (M-Z) type interferometer to implement the protocol. Construct a M-Z interferometer between Alice and Bob, both of input and output ports are in Alice's site, and only a portion of one arm in Bob's site. Bob still encodes his bits using the presence or not of PS, and Alice encodes secret bits with the two input modes of the M-Z interferometer. This version does not rely on the photon's polarization, so it reduces the requirement of the single-photon source.

In summary, we have proposed a new protocol for QKD based on quantum interference effect, in which Bob encodes his secret bits on the removing or inserting of a phase shifter. Because the protocol is sensitive to interference, the potential eavesdropper Eve is easily detected. Eve doesn't have full access to the entire quantum system, so the protocol is security according to the no-cloning principle of orthogonal states. In ideal cases, the efficiency of creating sifted key can theoretically reach 100\%, and the authorized parties are not required to check their bases during the process of creating sifted keys. All the devices of the implementation scheme are available currently, hence the protocol has the advantages of feasibility and high efficiency and may provide an interesting approach for quantum cryptography.

\begin{center}$\mathbf{Acknowledgments}$\end{center}

This work is supported by the National Natural Science Foundation
of China under Grant Nos. 61465013, 11264042, and 11465020; the
Program for Chun Miao Excellent Talents of Jilin Provincial
Department of Education under Grant No. 201316; and the Talent
Program of Yanbian University of China under Grant No. 950010001.


\begin{thebibliography}{999}
\bibitem{1} C.H. Bennett, G. Brassard, in Proceedings of IEEE International Conference on Computers, Systems, and Signal Processing, Bangalore, December 1984 (IEEE, New York, 1985), pp. 175-179.
\bibitem{2} A.K. Ekert, Phys. Rev. Lett. 67 (1991) 661.
\bibitem{3} C.H. Bennett, Phys. Rev. Lett. 68 (1992) 3121.
\bibitem{4} L. Goldenberg, L. Vaidman, Phys. Rev. Lett.75 (1995) 1239.
\bibitem{5} D. Bru{\ss}, Phys. Rev. Lett. 81 (1998) 3018.
\bibitem{6} H.-K. Lo, M. Curty, B. Qi, Phys. Rev. Lett. 108 (2012) 130503.
\bibitem{7} N. Gisin, G. Ribordy, W. Tittel, H. Zbinden, Rev. Mod. Phys. 74 (2002) 145.
\bibitem{8} V. Scarani, H. Bechmann-Pasquinucci, N.J. Cerf, M. Du\v{s}ek, N. L\"{u}tkenhaus, M. Peev, Rev. Mod. Phys. 81 (2009) 1301.
\bibitem{9} T.-G. Noh, Phys. Rev. Lett. 103 (2009) 230501.
\bibitem{10} Y. Sun, Q.-Y. Wen, Phys. Rev. A 82 (2010) 052318.
\bibitem{11} Z.-Q. Yin, H.-W. Li, W. Chen, Z.-F. Han, G.-C. Guo, Phys. Rev. A 82 (2010) 042335.
\bibitem{12} Z.-Q. Yin, H.-W. Li, Y. Yao, C.-M. Zhang, S. Wang, W. Chen, G.-C. Guo, Z.-F. Han, Phys. Rev. A 86 (2012) 022313.
\bibitem{13} M. Ren, G. Wu, E.Wu, H. Zeng, Laser Phys. 21 (2011) 755.
\bibitem{14} G. Brida, A. Cavanna, I.P. Degiovanni, M. Genovese, P. Traina, Laser Phys. Lett. 9 (2012) 247.
\bibitem{15} Y. Liu, L. Ju, X.-L. Liang, S.-B. Tang, G.-L.S. Tu, L. Zhou, C.-Z. Peng, K. Chen, T.-Y. Chen, Z.-B. Chen, J.-W. Pan, Phys. Rev. Lett. 109 (2012) 030501.
\bibitem{16} A.C. Elitzur, L. Vaidman, Found. Phys. 23 (1993) 987.
\bibitem{17} W.K. Wootters, W.H. Zurek, Nature (London) 299 (1982) 802.
\bibitem{18} T. Mor, Phys. Rev. Lett. 80 (1998) 3137.
\bibitem{19} J. Min\'{a}\v{r}, H.de Riedmatten, C. Simon, H. Zbinden, N. Gisin, Phys. Rev. A 77 (2008) 052325.
\bibitem{20} S.-B. Cho, T.-G. Noh, Opt. Express 17 (2009) 19027.
\bibitem{21} M.A. Hall, J.B. Altepeter, P. Kumar, Phys. Rev. Lett. 106 (2011) 053901.
\bibitem{22} P.C. Humphreys, B.J. Metcalf, J.B. Spring, M. Moore, X.M. Jin, M. Barbieri, W.S. Kolthammer, I.A. Walmsley, Phys. Rev. Lett. 111 (2013) 150501.
\end{thebibliography}
\end{document}